\documentclass[aps,prl,twocolumn,groupedaddress,showpacs,showkeys]{revtex4-1}
\usepackage{amsfonts}
\usepackage{mathtools}
\usepackage{amsmath}
\usepackage{amssymb}
\usepackage{float}
\usepackage{subdepth}
\usepackage{color}
\usepackage{graphicx}
\usepackage{natbib}
\usepackage{hyperref}
\usepackage{url}
\hypersetup{colorlinks, linkcolor={blue}, citecolor={blue}, urlcolor={blue}}

\begin{document}

\title{Spin transport and dynamic properties of two-dimensional spin-momentum locked states}
\author{Ping Tang$^{1,2}$}
\author{Xiufeng Han$^1$}
\author{Shufeng Zhang$^2$}
\email{zhangshu@email.arizona.edu}
\affiliation{$^1$ Beijing National Laboratory for Condensed Matter Physics, Institute of Physics, University of Chinese Academy of Sciences, Chinese Academy of Sciences, Beijing, 100190, China}
\affiliation{$^2$ Department of Physics, University of Arizona, Tucson, AZ 85721, USA}
\begin{abstract}

Materials with spin-momentum locked surface or interface states provide an interesting playground for
studying physics and application of charge-spin current conversion. To characterize their non-equilibrium magnetic and transport properties in the presence of a time-dependent external magnetic field and a spin injection from a contact, we
introduce three macroscopic variables: a vectorial helical magnetization, a scaler helical magnetization, and the conventional magnetization. We derive a set of closed dynamic equations for these variables by using the spinor Boltzmann approach with the collision terms
consistent with the symmetry of spin-momentum locked states. By solving the dynamic equations, we predict several intriguing magnetic and transport phenomena which are experimentally
accessible, including magnetic resonant response to an AC applied magnetic field, charge-spin conversion, and spin current
induced by the dynamics of helical magnetization.
\end{abstract}
\maketitle
\section{I. Introduction}
The electronic states with spin-momentum locking (SML) into mutually perpendicular directions occur at the surface of a topological insulator or at the interface with a strong Rashba interfacial spin-orbit coupling  \cite{RevModPhys.82.3045,manchon2015new,soumyanarayanan2016emergent}. These SML states are two-dimensional itinerant magnetic states without spontaneous magnetic moment in the absence of the magnetic field and there is no magnetic hysteresis response to the magnetic field. Yet, these materials have displayed profound magnetic phenomena. For example, an applied electric current can induce a non-equilibrium spin density known as the Edelstein effect (EE) \cite{edelstein1990spin,PhysRevLett.93.176601,kondou2016fermi,kondou2016fermi}, and reciprocally, a spin current injection produced by, e.g., spin pumping \cite{PhysRevLett.88.117601,PhysRevB.66.224403}, to such electronic surface states yields
an electric charge current termed as the inverse Edelstein effect (IEE) \cite{ganichev2002spin,PhysRevLett.112.096601,sanchez2013spin,PhysRevLett.113.196601,PhysRevLett.116.096602,zhang2015spin}. These observed EE and IEE phenomena have drawn considerable interest due to their potential applications for spin-based electronic devices \cite{RevModPhys.76.323}.

In this paper, we theoretically study the time-dependent magnetic and spin transport properties of such SML systems in the presence of the external spin injection and magnetic field. In the conventional magnetic materials, the time-dependent
magnetization dynamics is described by the Landau-Lifshitz-Gilbert equation with an additional Slonczewski spin transfer torque \cite{PhysRevB.54.9353,slonczewski1996current} when a spatially varying spin current is present. For the SML systems, there is no net magnetization without a magnetic field in the equilibrium even though the spin state of an electron with a given momentum is well-defined, i.e., the spin is ordered perpendicularly to the momentum. Such unique ordered spin configuration in the momentum space may be described by three macroscopic variables: a vectorial helical magnetization (VHM) $\boldsymbol{\xi} \equiv <\hat{\bf p}\times {\boldsymbol\sigma}>$ and a scalar helical magnetization (SHM) $\eta \equiv<\hat{\bf p}\cdot {\boldsymbol\sigma}>$, in addition to the conventional magnetization ${\boldsymbol m}\equiv<{\boldsymbol\sigma}>$, where $\hat{\bf p}={\bf p}/{\rm p}$ is the direction of momentum and ${\boldsymbol\sigma}$ is the Pauli matrix. Clearly, VHM and SHM characterize the relative orientation of the momentum and spin. In equilibrium, these macroscopic variables take simple values for the SML system, $\boldsymbol{\xi}_{\rm eq} \ne0$, $\eta_{\rm eq} =0$ and ${\boldsymbol m}_{\rm eq}=0$.

The main objective of the paper is to determine these variables in the presence of a driving force such as a time-dependent magnetic field or a spin current injected from a nearby contacting metallic layer, and more interestingly, to establish the relations between these variables and transport properties that can be measured experimentally. Due to the special band structure of SML systems, we find that the spin transport and dynamic properties display a number of unique characteristics and the above defined
three macroscopic variables provide a convenient way to understand and explain the non-equilibrium processes. This paper is organized as follows. In Sec.~II, we present our model and introduce a spinor form of the Boltzmann equation. In Sec.~III, we
derive the dynamic equations of these three macroscopic variables from the spinor Boltzmann equation, with appropriate
simplifications. In Sec.~IV, we solve the equations in several cases that are experimentally accessible. In Sec.~IV, we consider the effect of the time-dependent motion of the VHM on the spin pumping. Finally, we conclude the paper in Sec.~V.

\section{II. Model Hamiltonian and Spinor Boltzmann equation}

We consider a spin-orbit coupled two-dimensional band structure with a simple dispersion relation given by
\begin{equation}\label{Ha}
\hat{\varepsilon}_{\bf p}=\varepsilon_{\bf p}^{0}+\alpha ({\hat{\bf z} \times {\bf p} ) \cdot\boldsymbol{\sigma}}
\end{equation}
where $\varepsilon_{\bf p}^{0}$ is the spin-independent part of the electron dispersion which we will take zero for the surface
state of a topological insulator and
$\varepsilon_{\bf p}^{0}={\bf p}^2/2{ m^*}$ for a Rashba band, where $m^*$ is the effective mass, and $\alpha$ is the spin-orbit coupling constant. For a given momentum ${\bf p}$, the eigenvalues take $\varepsilon_{{\bf p}\pm}=\varepsilon_{\bf p}^{0} \pm \alpha{\rm p}$ with the corresponding spin eiegnstates $\chi_{{\bf p}\pm}$ satisfying $ (\boldsymbol{\Omega}_{\bf p}\cdot{\boldsymbol\sigma})\chi_{{\bf p}\pm}=\pm\chi_{{\bf p}\pm}$, where $\boldsymbol{\Omega}_{\bf p} \equiv (\hat{\bf z}\times \hat{\bf p})$ is the unit vector representing the quantization axis of the spin for a given momentum ${\bf p}$.
\begin{figure}
  \centering
  \includegraphics[width=8.6cm]{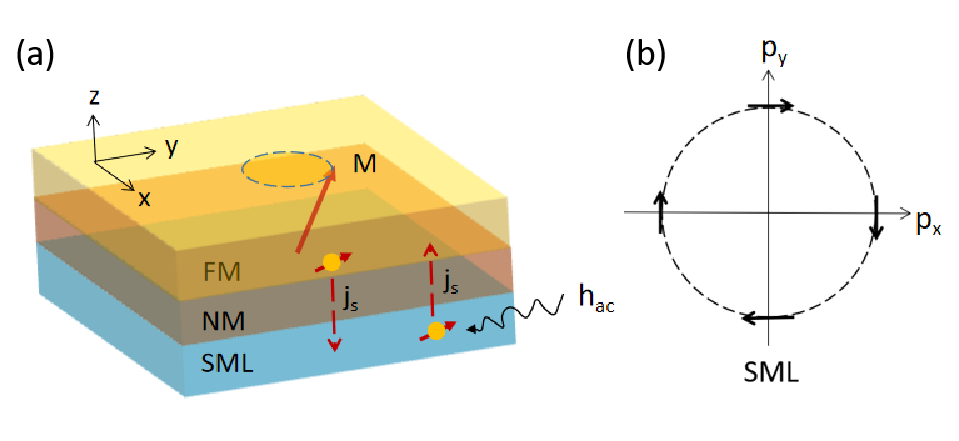}\\
  \caption{(a) The schematics of the SML surface in contact with a normal metal layer. On the other side of NM, a ferromagnetic
layer is served as a either spin injection source or spin detection probe with an attached normal metal. The role of the normal metal is to avoid the magnetic coupling between the spin source (the ferromagnetic layer) and SML layer while to allow spin current to flow through the entire layers. Inversely, the dynamic resonant states of SML excited by an AC magnetic field can pump a spin current into the normal metal that will be detected by the magnetic probe at the top. (b) A simple picture of spin directions at the Fermi circle of the SML states.}\label{1}
\end{figure}

The above model describes a simple one-electron SML state at equilibrium. We now turn on an AC magnetic field ${\bf h}_{ex} (t)$ and
a DC electric field ${\bf E}_{ex}$, in addition to a possible spin injection from a contacting normal metal (NM), as shown in Fig.~1.
The electron Hamiltonian with a given momentum then reads
\begin{equation}
\hat{H}_{\bf p}=\hat{\varepsilon}_{\bf p} + {\bf h}_{ex} (t) \cdot \boldsymbol{\sigma} + e{\bf E}_{ex} \cdot {\bf r}
\end{equation}
To calculate the spin transport properties in response to the magnetic and electric fields, we introduce a spinor form of the semiclassical Boltzmann distribution function for the SML band,
\begin{gather}\label{noneq}
\hat{f}({\bf p},t)=\hat{f}_0 ({\bf p})+f_c ({\bf p},t)+{\boldsymbol f}_s ({\bf p},t) \cdot{\boldsymbol{\sigma}}
\end{gather}
where $\hat{f}_{0}$ is the equilibrium Fermi distribution function and we take the distribution at $T=0$, i.e., $\hat{f}_0 ({\bf p})=\Theta(\varepsilon_F-\hat{\varepsilon}_{\bf p})$ with ${\varepsilon_F}$ the Fermi energy. In Eq. (\ref{noneq}), the non-equilibrium distribution function is separated into the spin-independent ($f_c$) and spin-dependent (${\boldsymbol f}_s$) parts. The generalized spinor Boltzmann equation is,
\begin{equation}\label{B}
  \frac{\partial\hat{f}}{\partial t} +e{\bf E}_{ex} \cdot\hat{\bf v}\left(-\frac{\partial\hat{f}_0}{\partial\hat{\varepsilon}_{\bf p}}\right)  + i\Big[\hat{H}_{\bf p}, \hat{f}\Big]=\left(\frac{\partial\hat{f}}{\partial t}\right)_{\rm col}
\end{equation}
where $\hat{\bf v} = \partial{\hat\varepsilon}_{\bf p}/{\partial {\bf p}}$ is the velocity and the anticommutator is implied for the product between Pauli matrices in this paper.

At this point, we want to emphasize that the presence of the commutator in Eq.~(\ref{B}) between $\hat{H}_{\bf p}$ and the spinor distribution function allows the electron to occupy the states that are {\em not} the spin eigenstates at equilibrium; this term represents the procession of the nonequilibrium electron spin around the effective magnetic field (the spin-orbit field $\alpha(\hat{\bf z}\times{\bf p})$ and the external magnetic field ${\bf h}_{ex}$). Thus, we do not assume that $\boldsymbol{f}_s$ is parallel to the $\boldsymbol\Omega_{\bf p}$. Recall that in a ferromagnetic metal, the spin dependent distribution function ${\boldsymbol f}_s$ is always confined to the states parallel or antiparallel to the local magnetization ${\bf M}$, i.e., ${\boldsymbol f}_s \propto {\bf M}$ and thus the Boltzmann equation
in the conventional ferromagnet has only two components (spin up and down relative to the local magnetization). The transverse component of spin accumulation or spin current injected to a ferromagnet is absorbed by the
ferromagnet within a few atomic distance due to the strong exchange interaction between the transverse spin and local magnetization; the absorption of the transverse spin current is considered as the manifestation of spin transfer torque. In the present case, we argue that the spin orbit coupling responsible for the dephasing of injected spins is much weaker compared to that of the transverse spin in conventional ferromagnets since a typical spin orbit coupling in a Rashba Hamiltonian or TI, which is about tens of meV \cite{zhang2015spin,kondou2016fermi,PhysRevLett.116.096602}, is much smaller than the exchange interaction in ferromagnets (of the order of eV); we will return to this point later when we model relaxation times. Thus,
we do not specify the direction of ${\boldsymbol f}_s$ and instead we will determine the dynamic equations relevant to it from the above spinor Boltzmann equation. The resulting commutator in Eq.~(\ref{B}) for an arbitrary ${\boldsymbol f}_s$ is
\begin{equation}\label{com}
\Big[\hat{H}_{\bf p}, \hat{f}\Big]= 2i\Big[ \Big(\alpha \hat{\bf z}\times {\bf p} + {\bf h}_{ex}\Big) \times {\boldsymbol f_s} \Big]
\cdot \boldsymbol{\sigma}
\end{equation}

The collision term on the right-hand side of Eq. (\ref{B}) has two contributions: one is the internal spin and momentum relaxations of the SML system, and the other is the interfacial scattering with the attached NM layer. We may parameterize these processes below,
\begin{align}\label{col}
  \left(\frac{\partial\hat{f}}{\partial t}\right)_{\rm col}=&-\frac{{f}_{c}+\boldsymbol f_L\cdot\boldsymbol{\sigma} }{\tau_p} - \frac{\boldsymbol{f}_T\cdot\boldsymbol\sigma}{\tau_{\phi} }\nonumber\\&+
\sum_{\bf p'}\hat{\Gamma}_{{\bf p}{\bf p}'}\left[ \hat{g}({\bf p}',t)-\hat{f}({\bf p},t) \right]
\end{align}
where we have introduced two characteristic relaxation times for the SML: the first term is caused by the momentum scattering of electrons, which is responsible for the relaxation of the longitudinal part of spin-dependent distribution function (relative to the spin-orbit field direction), i.e., $\boldsymbol f_L \equiv ({\boldsymbol f}_s \cdot {\boldsymbol\Omega_{\bf p}} ){\boldsymbol\Omega_{\bf p}}$ with a relaxation time $\tau_p$, and the second term is the spin dephasing for the transverse part of the spin-dependent distribution function $\boldsymbol f_T \equiv \boldsymbol f_s - ({\boldsymbol f}_s \cdot {\boldsymbol\Omega_{\bf p}} ){\boldsymbol\Omega_{\bf p}}$ with the relaxation time $\tau_{\phi}$. The last term in Eq.~(\ref{col}) represents the contribution from the interfacial scattering between the SML and NM layers with a transition rate $\hat{\Gamma}_{{\bf p}{\bf p}'}$ and $\hat{g}({\bf p}',t)$ being the spinor distribution function of the NM layer at the interface.

\section{III. Dynamic equations for macroscopic variables}

The Boltzmann equation, Eq.~(\ref{B}), along with Eqs.~(\ref{com}) and (\ref{col}), remains mathematically complicated since the collision terms
make the Boltzmann equation an integral equation. To further reduce the mathematical complication, in this Section, we propose to
generate the spin-diffusion-like equations from
the Boltzmann equation such that resulting simpler equations can be directly used for experimental analysis. The following
approximations are made. First, we assume the distribution function of the NM layer $\hat{g}({\bf p}',t)$ can be described by a time-independent single parameter, i.e., the spin chemical potential, defined as $\boldsymbol{\mu}_s=(1/e\mathcal{N}_F)\sum_{\bf p}{\rm Tr}_\sigma({\boldsymbol\sigma}\hat{g})$ with $\mathcal{N}_F$ its density of state at the Fermi level, which acts as a bias injecting the spin current into the SML layer. Note that the spin chemical potential in the NM layer may come from the spin
injection of the source layer such as the spin pumping of ferromagnetic layer, see Fig.~1. In principle, one should self-consistently determine the distribution function $\hat{g}({\bf p'},t)$ or the spin chemical potential $\boldsymbol{\mu}_s $; this will involve the Boltzmann equation and boundary conditions for $\hat{g}({\bf p}',t)$ at the interface of the NM and spin source layers. For the purpose of deriving the closed form of macroscopic dynamic equations, the presence of the spin chemical potential near the SML layer is sufficient. The second approximation is to assume that the electron transition across the interface conserves the spin and the interface transition rate
$\hat{\Gamma}_{{\bf p}{\bf p}'}$ is independent of the momentum ${\bf p}'$ of electrons in the NM layer; this is the assumption frequently used for modeling electron tunneling or diffusion across a rough interface. It is, however, that $\hat{\Gamma}_{{\bf p}{\bf p}'}$ does depend
on the momentum ${\bf p}$ of electrons in the SML layer even for the rough interface since the electron spin in the SML layer is coupled to its momentum. The symmetry of the SML states demands $\hat{\Gamma}_{{\bf p}{\bf p}'}$ to have the following form,
\begin{equation}\label{Gamma}
  \hat{\Gamma}_{{\bf p}{\bf p}'}=\left( \frac{1}{\tau_c} \right)+ \left( \frac{1}{\tau_s} \right) \boldsymbol\Omega_{\bf p}\cdot{\boldsymbol\sigma}
\end{equation}
where $1/\tau_c$ and $1/\tau_s$ characterize the sum and difference of the transition rate across the interface for two spin subbands $``\pm"$ of the SML, respectively. In the case of a single subband at the Fermi Level, i.e., a topological insulator band, $1/\tau_c =1/\tau_s$ \cite{PhysRevB.94.184423}, while in the case of a Rashba band we assume that the transition rates for two subbands are the same in the limit $\varepsilon_F\gg\alpha{\rm p}_F$ and thus $1/\tau_s=0$.

Finally, we assume the momentum relaxation time is much faster than the transverse spin relaxation time ($\tau_p/\tau_{\phi}\ll 1$), as we have discussed above. Physically, $\tau_p$ is due to the impurity or defect scattering, while $\tau_{\phi} $ involves inelastic or interband scattering. More quantitatively, $\tau_{\phi}$ scales inversely
with the strength of spin orbit coupling which is the order of several tens of meV for the TI or Rashba systems, and thus $\tau_{\phi}$ is
about picoseconds while the momentum relaxation time is typically a few femtoseconds \cite{PhysRevLett.112.096601}. The separation between these two time scales are critically important since we are able to treat the dynamics for the longitudinal and transverse spins differently. In fact, we will limit our dynamic description between these two time scales such that the longitudinal spin reaches steady states instantly. Equivalently, we take the limit $\partial f_c/\partial t =0 $ and
$\partial\boldsymbol{f}_L / {\partial t} =0$ in the Boltzmann equation, and we focus the time dependence of the
distribution function on the $``slow"$ dynamics of the transverse spin part $\boldsymbol f_T$.

With above simplifications, we can now explicitly establish the dynamic equations for three macroscopic variables by inserting the Boltzmann equation Eq.~(\ref{B}) along with the explicit relations of Eqs.~(\ref{com}), (\ref{col}) and (\ref{Gamma}) into the definitions of the VHM, $\boldsymbol{\xi} (t) =
\sum_{\bf p}  {\rm Tr}_{\sigma} (\hat{\bf p}\times \boldsymbol{\sigma} \hat{f} ),$ the SHM, $\eta (t)= \sum_{\bf p}  {\rm Tr}_{\sigma} (\hat{\bf p}\cdot \boldsymbol{\sigma} \hat{f} ),$ and the conventional magnetization, ${\boldsymbol m}(t) =\sum_{\bf p}  {\rm Tr}_{\sigma} (\boldsymbol{\sigma} \hat{f} )$. After a tedious but straightforward algebra, we obtain,
\begin{align}\label{Eq}
\frac{\partial{\boldsymbol m}}{\partial t}=&\omega_0\Big[\hat{\bf z}
\times\boldsymbol\xi-\eta\hat{\bf z}\Big]+2\boldsymbol{h}_{ex}\times{\boldsymbol m}-\frac{{\boldsymbol m}-{\boldsymbol m}_{\rm p}}{\tau}\nonumber\\&+g_{\rm int} \Big[ \big(\boldsymbol\mu_s\cdot\hat{\bf z}\big)\hat{\bf z}+\frac{1}{2}\big(\hat{\bf z}\times\boldsymbol\mu_s\big)\times\hat{\bf z} \Big] \\
\frac{\partial\boldsymbol\xi}{\partial t}=&\omega_0 \hat{\bf z}\times\Big[{\boldsymbol m}-{\boldsymbol m}_{\rm p}\Big]+2\eta\boldsymbol h_{ex}-\frac{\boldsymbol\xi-\boldsymbol\xi_{eq}}{\tau}\\
 \frac{\partial\eta}{\partial t}=&\omega_0 \hat{\bf z}\cdot{\boldsymbol m}-2\boldsymbol h_{ex}\cdot\boldsymbol\xi-\frac{\eta}{\tau}
\end{align}
where $\omega_0= 2\alpha {\rm p}_{F}$ with ${\rm p}_F$ the Fermi momentum, $g_{\rm int}=e\mathcal{N}_F/\tau_c$ is an interface spin conductance, $1/\tau=1/\tau_\phi+ 1/\tau_c $ and $\boldsymbol m_{\rm p}=\sum_{\bf p} {\rm Tr}_{\sigma} [(\boldsymbol{f}_{\boldsymbol L}\cdot\boldsymbol\sigma)\boldsymbol{\sigma}]$ reads
\begin{align}\label{mp}
{\boldsymbol m}_{\rm p}=&\frac{e\mathcal{N}_F\tau_p}{2(\tau_p+\tau_c)}\hat{\bf z}\times(\boldsymbol{\mu}_s\times\hat{\bf z})+\tau_cq_{EE}\hat{\bf z}\times {\bf E}_{ex}
\end{align}
where $q_{EE}$ is the EE coefficient,
\begin{equation}\label{ee}
q_{EE} =\begin{cases}
  \frac{\alpha e \tau_p m^*}{2\pi(\tau_p+\tau_c)} &\mbox{for Rashba,}\\
  -{{\rm sgn}(\varepsilon_F)}\frac{e{\rm p}_F\tau_p}{4\pi(\tau_p+\tau_c)} &\mbox{for TI}
   \end{cases}
\end{equation}
where ${\rm sgn}(\varepsilon_F)=+1$ for $\varepsilon_F>0$ and ${\rm sgn}(\varepsilon_F)=-1$ for $\varepsilon_F<0$. Eq. (8)-(10) along with Eq. (11) are our main results and they can be broadly used for capturing the dynamics of the SML states in the presence of external fields and spin current injection. We shall point out that the difference between the total
magnetization ${\boldsymbol m}$ and the ${\boldsymbol m}_{\rm p}$: the latter
describes the shift of momentum center due to the presence of an electric field and spin current injection. Since we assume a fast relaxation for the electron
momentum, ${\boldsymbol m}_{\rm p}$ is treated as a steady state solution. Microscopically, ${\boldsymbol m}_{\rm p}$ is comprised of the states
with the longitudinal spins (relative to the spin-orbit filed direction ${\boldsymbol\Omega_{\bf p}}$), while the ${\boldsymbol m}$ includes
both transverse and longitudinal spin components. It is the transverse component of ${\boldsymbol m}$ that gives arise to the time-dependent motion. When a spin current injected from the contact, its longitudinal spin component induces a spin accumulation ($\boldsymbol m_{\rm p}$) and thereby converts to an electric
current, i.e., the inverse Edelstein effect, while its transverse spin component is equivalent to the spin transfer torque which drives the magnetization dynamics. One might compare this picture with the conventional spin injection to a ferromagnet where the transverse spin current
leads to the spin transfer torque on a macros-spin while the longitudinal component generates a magnetoresistance (or spin accumulation).

\section{IV. Applications of the Dynamic Equations}

These equations may be considered as an extension of the Landau-Lifshitz-Gilbert-Slonzcewski (LLGS) to the SML systems. For a
conventional itinerant ferromagnet, the LLG equation involves only one macroscopic variable in magnetization; we have three coupled equations for three helix-dependent magnetizations $\boldsymbol{\xi} (t)$, ${\boldsymbol m} (t)$ and $\eta (t)$. Therefore, it may be more appropriate to compare Eq.~(8)-(10) with the LLGS for antiferromagnets in which the dynamics of the staggered magnetic moment is always coupled to the magnetization because a time-dependent change of the staggered moment is only possible when the magnetic moment of each sublattice does not exactly compensate for each other. In the present case, our VHM and SHM are coupled to the in-plane and out-of-plane direction of the conventional magnetization through the spin-orbit coupling, respectively. In this Section, we solve these equations in two simple cases that can be readily tested experimentally.

\subsection{A. Magnetic resonance}

As in the cases of ferromagnets and antiferromagnets, the dynamic equations contain characteristic resonant states. To see this, we ignore all the relaxation terms and then take simple time
dependent solutions, i.e., $\boldsymbol\xi(t)$, $\boldsymbol m(t)$ and $\eta(t)\propto {\rm exp}(-i\omega t)$. By placing them into Eqs.(8)-(10), we immediately obtain three degenerate resonant modes at frequency $\omega = \omega_0$. There are two transverse modes consisting of the left-handed and right-handed precessions of the VHM and in-plane component of magnetization, and one longitudinal mode representing the oscillation of SHM and longitudinal component of magnetization. In Fig. 2, we show the corresponding microscopic spin dynamics in momentum space for the three modes. We shall emphasize that the resonant frequency is controlled by the spin-orbit coupling or the spin-momentum locking parameter $\alpha$; this is rather different from the conventional magnetic systems where the resonant frequency is determined either by the anisotropy field for a ferromagnet, or by the geometric mean of the exchange coupling and the anisotropy field for an antiferromagnet.
\begin{figure}
  \centering
  \includegraphics[width=7.6cm]{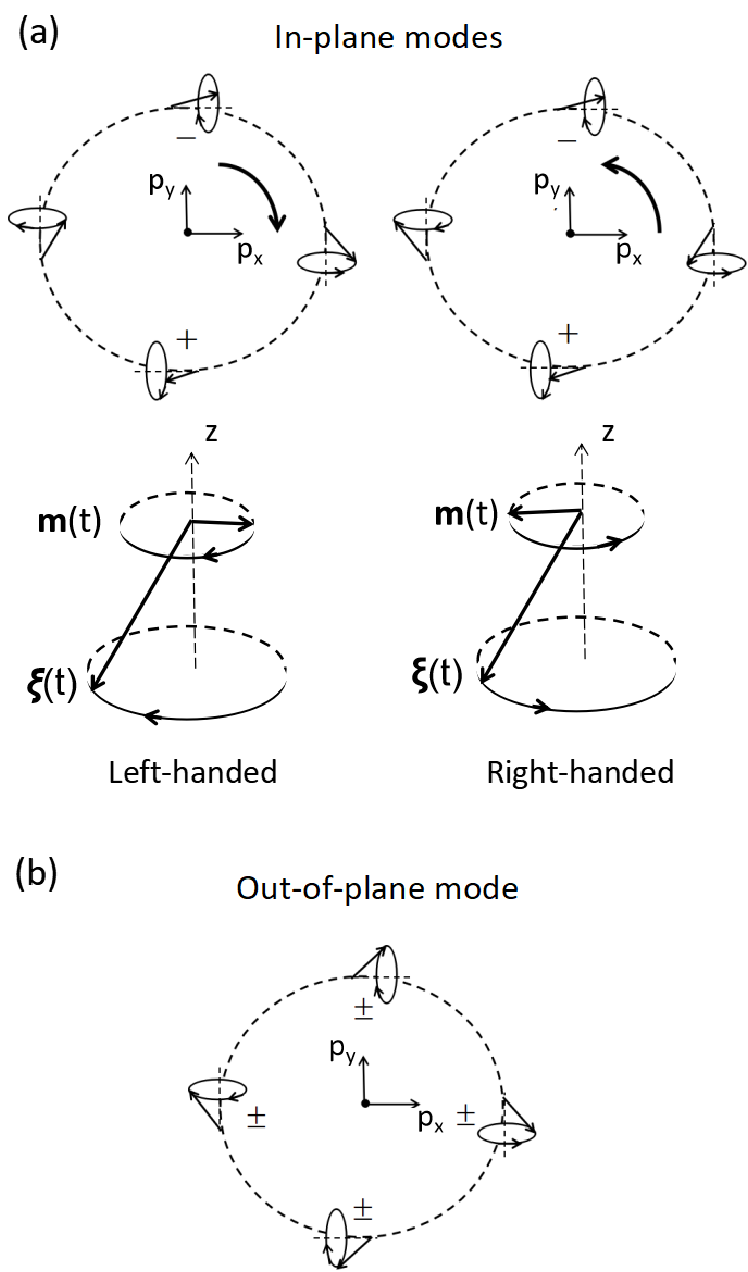}\\
  \caption{Microscopic dynamics of the electron spin in momentum space for three degenerate resonant modes, in which each spin precesses along the local spin-orbit field. (a) The two in-plane modes include the right-handed and left-handed precession of mutually coupled VHM $\boldsymbol\xi$ and the in-plane component of ${\boldsymbol m}$. (b) The out-of-plane mode represents the coupled oscillation of SHM $\eta$ and the out-of-plane component of ${\boldsymbol m}$. The small black arrow represents the electron spin and ``$\pm$" is the sign of spin projection along z direction.}\label{2}
\end{figure}

The above resonant states can be excited by applying a time-dependent magnetic field, similar to the ferromagnetic resonance (FMR). To excite an in-plane resonant mode, one applies a combined out-of plane DC and a small in-plane AC magnetic field ${\bf h}_{ex} (t) ={h_0} \hat{\bf z} + \delta{h}_{\rm ac}(\cos\omega t\hat{\bf x}+\sin\omega t\hat{\bf y})$. By placing the magnetic field into Eq.(8)-(10) with no electric field ${\bf E}_{ex}=0$ and spin injection ${\boldsymbol\mu}_s=0$ and by defining the in-plane VHM susceptibility, $\delta\xi=\chi_{\rm in}\delta h_{\rm ac} e^{-i\omega t}$, where $\delta\xi=\delta\xi_x-i\xi_y$, we obtain
\begin{equation}
\chi_{\rm in}(\omega)=\left(\frac{4h_0\xi_{\rm eq}}{1+\omega_0^2\tau^2}\right)\frac{(1-i\omega\tau)}{\omega^2-\omega_0^2-2h_0\omega+2i\omega/\tau}
\end{equation}
where we ignored the second-order term in $1/\tau$ and $h_0$, and the magnetization is related to the VHM through
\begin{equation}
{m} =\frac{-i\omega_0\tau}{1-i\omega\tau+2ih_0\tau}\delta\xi
\end{equation}
where $m=m_x-im_y$. Note that the DC magnetic field ${h_0}$ is necessary
to excite the in-plane resonant mode since the resonant amplitude of VHM is proportional to $h_0$. Clearly, as shown in Eq.~(13), the DC magnetic field applied along $\hat{\bf z}$ direction can also lift the degeneracy of two in-plane processional modes, i.e., $\omega_{\pm}=h_0\pm\omega_0$, where $+$$(-)$ correspond to the right (left)-handed precessions.

Similarly, if one applies a small AC field along the $\hat{\bf z} $ axis, ${\bf h}_{ex}(t) =h_{ac}e^{-i\omega t}\hat{\bf z}$, one is able to excite the dynamics of SHM $\eta (t)$. The out-of-plane susceptibility of the SHM $\chi_{\rm out}(\omega)\equiv \eta/h_{ac}$ reads,
\begin{equation}\label{l}
\chi_{\rm out}(\omega)=\frac{2(1/\tau-i\omega)\xi_{\rm eq}}{\omega^2-\omega_0^2+i\epsilon}
\end{equation}
with the magnetization in this case
\begin{equation}
{m}_z =-\frac{\omega_0\tau}{1-i\omega\tau}\eta
\end{equation}
In contrast to the in-plane modes, the out-of plane mode can be directly by an AC magnetic field along the z direction.

\subsection{B. Effects of spin injection}

As we mentioned earlier, the spin current injection from a contacting conductor has two effects. One is to accumulate spins in
the SML states, and consequently the spin accumulation leads to a charge current via spin orbit coupling, or the IEE effect. The other effect is a spin transfer torque exerted on the magnetization if the spin current is not parallel to the direction of local spin-orbit field $(\boldsymbol\Omega_{\bf p})$. To quantitatively include both effects, we set ${\bf h}_{ex}=0$ and first consider the steady state solution, i.e., $\partial{\boldsymbol{\xi}}/\partial t = \partial\eta/\partial t=\partial{\boldsymbol m}/\partial t =0$. From Eq.~(8)-(10), we have
\begin{align}
 {\boldsymbol m}=&{\boldsymbol m}_{\rm p}+ \frac{ \tau g_{\rm int}}{1+\omega_0^2 \tau^2}\Big[ \big(\boldsymbol\mu_s\cdot\hat{\bf z}\big)\hat{\bf z}+\frac{1}{2}\big(\hat{\bf z}\times\boldsymbol\mu_s\big)\times\hat{\bf z} \Big], \\
 {\delta\boldsymbol{\xi}}=&\frac{\omega_0 \tau^2g_{\rm int}}{2(1+\omega_0^2 \tau^2)}\hat{\bf z}\times{\boldsymbol\mu_s},\\
\eta=&\frac{\omega_0\tau^2g_{\rm int}}{1+\omega_0^2 \tau^2 }\hat{\bf z} \cdot\boldsymbol\mu_s.
\end{align}
The deviation of the VHM and SHM from the equilibrium values implies that the spin and momentum are no longer locked into completely perpendicular directions due to the spin transfer torque; this leads to an additional magnetization beyond the simple momentum-shift-relevant magnetization (${\boldsymbol m}_{\rm p}$), i.e., the second term in Eq.(17).

The spin current injected into the SML layer is correlated with the spin chemical potential at the interface. In the steady state, the
spin current across the interface is given by,
\begin{align}\label{js}
  \boldsymbol{j}_s=&\sum_{{\bf p}{\bf p'}}{\rm Tr}_\sigma\left\lbrace  \boldsymbol\sigma \hat{\Gamma}_{{\bf p}{\bf p}'} \left[ \hat{g}({\bf p}')-\hat{f}({\bf p})\right] \right\rbrace\nonumber\\
  =&\frac{e\mathcal{N}_F}{2(\tau_p+\tau_c)}\hat{\bf z}\times(\boldsymbol\mu_s\times\hat{\bf z})-q_{EE}\hat{\bf z}\times {\bf E}_{ex}\nonumber\\&+\Big(1-\zeta\Big) g_{\rm int}\Big[ \big(\boldsymbol\mu_s\cdot\hat{\bf z}\big)\hat{\bf z}+\frac{1}{2}\big(\hat{\bf z}\times\boldsymbol\mu_s\big)\times\hat{\bf z} \Big]
\end{align}
where the first term is the conventional spin current and the second term is the electric current-driven EE effect, and the last term represents the spin transfer torque from the spin injection whose spin component is not parallel to $({\bf \Omega_p})$ with a backflow factor $\zeta=\tau\tau_c^{-1} (1+ \omega_0^2 \tau^2)^{-1}$. Similarly, the electric current in the SML layer also gains a contribution from the spin transfer torque,
\begin{align}\label{je}
\boldsymbol{j}_e
=\sigma_e{\bf E}_{ex}+ \left( \lambda_{IEE}+\frac{ e\tau\omega_0 g_{\rm int}}{1+\omega_0^2 \tau^2}
\right) \hat{\bf z}\times\boldsymbol\mu_s
\end{align}
where $\sigma_e$ is the electric conductivity and $\lambda_{IEE}=-(2e/\hbar) q_{EE}$ is the conventional IEE coefficient. Besides the IEE term, there is an additional contribution to the electric current, which is irrelevant to the momentum shift and caused by the spin torque. Note that here we choose the electric field and spin accumulation in the NM as the driving forces and hence there exists an Onsager reciprocal relation $-(2e/\hbar)q_{EE}=\lambda_{IEE}$.

\subsection{C. Spin pumping}

Spin pumping is the reciprocal effect of spin current injection, similar to the non-magnetic metal-ferromagnet bilayer system, in which the time-dependent magnetization generates an outgoing spin current to the contacting non-magnetic metal. In this Section,
we formulate the spin pumping current due to the dynamics of the VHM and SHM. The pumping spin current across the interface can
be defined as,
\begin{align}\label{js}
  \boldsymbol{j}_s^{\rm pump}=\sum_{{\bf p}{\bf p}'}{\rm Tr}_\sigma\lbrace{\boldsymbol\sigma}\hat{\Gamma}_{{\bf p}{\bf p}'}\hat{f}({\bf p},t)\rbrace
\end{align}
where we do not include the ``flow back" of the spin current by the induced spin accumulation in the contact layer since it is the second effect of dynamics of VHM and SHM. By utilization Eq.~(3) and (7), one can immediately identify that the spin pumping in Eq.~(22) is proportional to the total magnetization ${\bf m}(t)$. Provided that the SML is in resonant modes and expressing the magnetization in terms of the time-derivatives of VHM and SHM, we find
\begin{align}
 \boldsymbol{j}_s^{\rm pump} =&-\frac{1}{\tau_c\omega_0} \left\lbrace
 {\hat{\bf z}}\times \frac{\partial \boldsymbol{\xi}}{\partial t}-\frac{1}{\omega_0 \tau}\hat{\bf z}\times\left(\frac{\partial\boldsymbol\xi}{\partial t}\times\hat{\bf z}\right)\right\rbrace\nonumber\\
&+\frac{1}{\tau_c\omega_0} \left( 1-\frac{i}{\omega_0 \tau} \right) \frac{\partial \eta}{\partial t}\hat{\bf z}
\end{align}
where the dynamics of VHM and SHM pump out an in-plane and an out-of-plane polarized spin current, respectively.

We want to emphasize the differences between the conventional spin pumping in ferromagnets and the above formula. In our
model, the dynamics of the VHM and SHM are induced by the AC magnetic field and we have used the approximation which
is valid up to the first-order in the magnetic field. Thus, the spin current pumping contains only the AC component. In the
conventional pumping, the magnetization dynamics could be generated by various methods and the the spin pumping
formula is written beyond the linear response. As a result, the conventional spin pumping contains both AC spin current
and a higher order DC component of the spin current.

\section{V. Discussion and summary }

We have considered the magnetic and spin transport properties in the presence of the time-dependent magnetic field and
spin current injection from a contact. We want to comment on the differences of the SML spin transport compared to other materials.

Up until now, we have not included any interaction among spins, and thus it is more appropriate to classify the SML as
a paramagnetic materials with a momentum dependent magnetic field on each particle. The magnetic resonant frequency is given by
the strength of spin-orbit magnetic field $\alpha{\rm p}_F$, similar to the external magnetic field as the paramagnetic resonance. On the other hand, the spins of electrons of the
SML is perfectly ordered in a helical state in the momentum space, similar to the ferromagnetic or antiferromagnetic spins ordered in real space. Although the spin ordering in the SML is not driven by the exchange interaction, there are some shared spin transport
properties such as spin dephasing and spin pumping.

In conventional ferromagnetic systems, there are two different degree of freedom for the magnetization dynamics and spin
transport. The magnetization density consists all electrons of occupied states while the spin transport is confined to the electron
near or at the Fermi level. Thus the dynamics involves the time-dependence of the magnetization and of the conduction
electrons. The magnetization dynamics are considered much slower than that of the conduction electrons, even though they
are strongly coupled. In the SML system, both magnetization and transport are governed by the states near the Fermi level.

A key observation of the SML is the different spin relaxations for the longitudinal and transverse components.
For conventional ferromagnetic metals, the spins have much shorter transverse relaxation time (or dephasing time) than the
longitudinal spins, and thus the magnetization dynamics modeled by the LLGS do not address the conduction electrons, but the
much slow dynamics of the local magnetization. While for the SML system, there is no spontaneous local magnetization. On the other
hand, the transverse spins relax much slower than the longitudinal spins, and thus our dynamic equations address the
dynamics of the conduction electrons.

The VHM and SHM provide a useful tool to visualize the spin orientation of the SML. When the VHM deviates from the ground state,
the in-plane component of the VHM indicates the degree of the spin tilting away from the perfect perpendicular locking between
the momentum of the spin.

In summary, we introduce three macroscopic variables for the SML and establish their equations of motion. Among other things, we have discussed the magnetic resonant frequency, spin injections associated with the EE and IEE, and we propose a spin pumping
formalism.

\section{VI. Acknowledges }
This work was partially supported by US National Science Foundation under Grant No. ECCS-1708180, the
National Natural Science Foundation of China(NSFC, Grant No.11434014) and China Scholarship Council (CSC, [2017]3109).
\bibliography{Refer}
\end{document}